\documentclass[runningheads]{llncs}
\usepackage{amsfonts}
\usepackage{hyperref}
\usepackage{graphicx}

\usepackage{floatrow}
\DeclareFloatFont{tiny}{\scriptsize}% "scriptsize" is defined by floatrow, "tiny" not
\newcommand{\gitlink}[0]{\url{https://github.com/DocPierro/optimised_spd}}

\usepackage{algpseudocode,algorithm,algorithmicx}

\usepackage{booktabs}
\usepackage{amsmath}
\usepackage{tikz}
\usetikzlibrary{automata,shapes.multipart} 
\usetikzlibrary{arrows,petri}
\usetikzlibrary{arrows}
\usetikzlibrary{shapes,backgrounds}
\usetikzlibrary{arrows,shapes,backgrounds,patterns,decorations.pathreplacing,decorations.pathmorphing,decorations.markings,shadows,shapes.misc,calc,positioning}
\usetikzlibrary {graphs}

\usepackage[colorinlistoftodos]{todonotes}
\definecolor{forestgreen}{rgb}{0.13,0.55,0.13}
\definecolor{amber}{rgb}{1.0,0.49,0}
\definecolor{cargreen}{rgb}{0.0,0.8,0.6}
\definecolor{aureolin}{rgb}{0.54,0.17,0.89}
\definecolor{bured}{rgb}{0.8,0.0,0.0}

\colorlet{orange}{red!70!yellow}
\colorlet{vert}{green!70!blue}
\colorlet{mauve}{blue!70!red}
\colorlet{brouge}{red!70!blue}
\colorlet{frouge}{red!40!black}

\usepackage[T1]{fontenc}
\usepackage{graphicx}
\usepackage{pgfplots}
\usepackage{subcaption}

\begin{document}
%
%\title{Stochastic process discovery via maximum likelihood optimisation}
\title{A framework for optimisation based stochastic process discovery\thanks{Accepted for publication at International Conference on Quantitative Evaluation of SysTems (QEST) 2024.}
}

\author{
Pierre Cry\inst{1}%\orcidID{1111-2222-3333-4444} 
\and
Andr\'as Horv\'ath\inst{2}%\orcidID{2222--3333-4444-5555}
\and
Paolo Ballarini\inst{1}%\orcidID{0000-1111-2222-3333} 
\and
Pascale Le Gall\inst{1}
}
\institute{Université Paris Saclay, CentraleSupélec, Gif-sur-Yvette, France 
\email{\{pierre.cry,paolo.ballarini,pascale.legall\}@centralesupelec.fr}\\
\and 
Università di Torino, Torino, Italy\\
\email{horvath@di.unito.it}}
\maketitle              % typeset the header of the contribution
\begin{abstract}
Process mining is concerned with deriving formal models capable of reproducing the behaviour of a given organisational process by analysing observed executions collected in an \emph{event log}. The elements of an event log are finite sequences (i.e., \emph{traces} or \emph{words}) of actions. 
Many effective algorithms have been introduced which issue a \emph{control flow} model (commonly in Petri net form) aimed at reproducing, as precisely as possible, the \emph{language} of the considered event log. However, given that identical executions can be observed several times, traces of an event log are associated with a \emph{frequency} and, hence, an event log inherently yields also a \emph{stochastic language}.
By exploiting the trace frequencies contained in the event log, the stochastic extension of process mining, therefore, consists in deriving stochastic (Petri nets) models capable of reproducing the likelihood of the observed executions. In this paper, we introduce a novel stochastic process mining approach. Starting from a ``standard'' Petri net model mined through classical mining algorithms,  we employ optimization to identify optimal weights for the transitions of the mined net so that the stochastic language issued by the stochastic interpretation of the mined net closely resembles that of the event log. The optimization is either based on the maximum likelihood principle or on the earth moving distance. Experiments on some popular real system logs show an improved accuracy w.r.t. to alternative approaches. 

\keywords{Stochastic process mining  \and Stochastic Petri nets \and Maximum-likelihood} \and Earth Movers Distance \and Weights estimation
\end{abstract}
\section{Introduction}
\label{sec:intro}
%Process mining~\cite{10.5555/2948762}, inductive miner~\cite{DBLP:conf/apn/LeemansFA13}, alpha~\cite{1316839}, heursitic~\cite{tonbeta166}, linear programming~\cite{10.1007/978-3-540-68746-7_24}, fodina~\cite{VANDENBROUCKE2017109}, split~\cite{10.1007/s10115-018-1214-x}

A \emph{business process} consists in a set of activities performed so to accomplish a specific organisational goal. 
Understanding business processes is of prime importance for organisations in many respects: firstly, to obtain a formal description of often undocumented processes, then to use the obtained \emph{process model} to monitor and possibly improve the process under study. 
The principle goal of \emph{process mining}~\cite{10.5555/2948762}  is to   \emph{discover} a process model by exploiting the data that information systems of organisations accumulate in the form of \emph{event logs}. An event log consists of a collection of traces where each \emph{trace} is formed by a sequence of timestamped activities representing one observed case of the process. Since the same sequence of actions may be observed in multiple cases of a process, the same event log may contain multiple occurrences of the same trace, therefore leading to  \emph{trace frequency} being inherently associated with traces of a log: trace frequency quantifies how many times each unique sequence of actions has been observed, hence how likely it is the process exhibits the corresponding behavior.

\emph{Process discovery} algorithms~\cite{DBLP:conf/apn/LeemansFA13,1316839,tonbeta166,10.1007/978-3-540-68746-7_24,VANDENBROUCKE2017109,10.1007/s10115-018-1214-x} analyse an event log in order to \emph{mine} a formal model, commonly a Petri net,  aimed at capturing the \emph{control-flow}  of the observed process and hence reproducing the \emph{log language}, that is: the \emph{model language}, issued by the model's operational semantics, should be close to the log language. 
% given by the sequence of events that can occur in the model, should be close to  the log language. 
Evaluating the resemblance between a mined model and its corresponding log is the subject of \emph{conformance checking} analysis, which includes criteria such as assessing how many traces of the log are reproduced by the model  (\emph{fitness}) and conversely, how many traces of the model are not amongst that of the log (\emph{precision}). Certain mining algorithms, e.g., inductive miner~\cite{DBLP:conf/apn/LeemansFA13}, guarantee, by construction,  that the mined model fits the log.
%\plg{PLG: Here, "fits" has to be understood up the notion of model fitness ?}\pb{Yes precisely}
%Process mining approaches have proven fundamental  to improve business process management and have  attracted the interest of a wide research community. Nevertheless `standard'' \emph{discovery} approaches  are affected by an inherent weakness as they do not consider a very valuable source of information contained in the logs, that is,  trace frequency. This may  severely undermine the ability to take business analysis 

Although proven effective in many respects, "standard" \emph{discovery} approaches are affected by an inherent weakness as they do not consider a very valuable source of information contained in the logs, that is,  trace frequency. This may severely undermine the ability to improve the modelled business as the discovered models cannot reproduce the likelihood of the observed behaviours.  That leads to \emph{stochastic process mining} whose goal is discovering stochastic models which, other than the control flow also reproduce the likelihood of the observed process. 
This can be achieved either \emph{indirectly}, i.e., using a standard miner to obtain a control flow model and then, based on the trace frequencies, devising adequate \emph{weights} to convert it into a stochastic model~\cite{DBLP:conf/icpm/BurkeLW20}, or \emph{directly}, i.e., by devising a stochastic model straight from the log~\cite{DBLP:conf/apn/BurkeLW21}. 
In the stochastic settings, the resemblance between models and logs (i.e.,  \emph{stochastic conformance checking}) concerns the resemblance between \emph{stochastic languages}.  The most popular Earth Movers Stochastic Conformance (EMSC) measure~\cite{10.1007/978-3-030-26643-1_8,LEEMANS2021101724} is obtained as an adaptation of the Earth Movers Distance (EMD or Wasserstein distance) to the stochastic languages case.

In this paper we consider the indirect stochastic process discovery approach and
propose a framework based on numerical optimisation to obtain optimal weights for the transitions of the Petri net. Our approach relies on calculating the probability that a given trace is produced by a stochastic Petri net as function of its transition weights. We do so by the construction of a dedicated \emph{unfold directed acyclic graph} whose nodes store the probability of finite prefixes as they are unfolded from the Petri net's reachability graph. 
Then, in order to optimize the weights, we need to measure the distance between the stochastic language of the event log and that of the Petri net.  We consider two distance measures, the Kullback-Leibler divergence and the earth moving distance.
Minimizing the Kullback-Leibler divergence corresponds to the maximum likelihood principle and leads to an efficient way to measure the divergence between the event log and the Petri net model. The earth moving distance is more popular in the area of stochastic languages but requires considerably more computational effort. To the best of our knowledge divergence based numerical weight optimization has not been experimented with before.

The paper is organised as follows: in Section~\ref{sec:prelim} we introduce preliminary notions the remainder of the manuscript relies upon; in Section~\ref{sec:method}, we describe the novel stochastic discovery approach we propose whereas in Section~\ref{sec:results} we demonstrate its application through a number of experiments on popular real-life event logs. We wrap up the manuscript with conclusive remarks and future perspectives in Section~\ref{sec:conclusion}.

%\medskip
%\noindent
%\pb{add reference to Conformance checking literature ?}\\
%\pb{certainly add reference to entropy based stochastic Conformance checking literature }\\
%\pb{add reference to Rogge-Stolte method: in the Related work subsection?}

\subsection{Related work}
    Not many methods have been proposed so far in the literature to discover  stochastic models from an event log. In Rogge-Solti \emph{et al.}'s seminal work  ~\cite{5f0e8dd04572478fb450eefe4211d69f} authors introduced a framework to discover generalised stochastic Petri net (GSPN) models extended with generally distributed timed transitions so to allow for performance analysis of the mined process. 

     Burke \emph{et al.}'s~\cite{DBLP:conf/icpm/BurkeLW20} instead address the problem of  converting a workflow Petri net, mined through a conventional  discovery algorithm, into an adequate stochastic workflow Petri net through  weight estimation. Notice that, differently from ~\cite{5f0e8dd04572478fb450eefe4211d69f}, the class of stochastic Petri net considered in this case is the subclass of GSPN whose transitions are solely \emph{immediate}, i.e., time delays are not captured by the model (hence the timestamps of the event log are ignored), only the probability of events is accounted for.   
     Specifically \cite{DBLP:conf/icpm/BurkeLW20} introduced six  weight estimators, that combine summary statistics computed on the log (e.g.,  number of times  a subsequence of activities  is observed on  the log's  traces) with statistics computed on the model and taking into account  structural relationships between the Petri net nodes (e.g.,  \emph{transitions causality}). 
     These estimators enjoy being computationally light, as, by definition, they do not need assessing the language of the Petri net model. However, the price for such simplicity is paid in terms of conformance as the distance between the resulting model's and the log's stochastic languages appears, in many cases, to be far from optimal. 
     In a follow up work~\cite{DBLP:conf/apn/BurkeLW21} the same authors introduced a framework to directly discover an \emph{untimed} GSPN model from a log based on traces' frequency.

\section{Preliminaries}
\label{sec:prelim}
We introduce a number of notions/notations that will be used in the remainder. 

\noindent
\textit{Alphabets, traces,  languages}.
%In order to characterise event logs as well as the language of a model (mined from a log) 
We let  $\Sigma$ denotes the alphabet of an event log's activities (we use  letters to denote activities of an alphabet, e.g., $\Sigma=\{a,b,c\}$) and  $\Sigma^*$  the set of possibly infinite traces (words) composed of activities in $\Sigma$ where $\epsilon\in\Sigma^*$ represents the empty trace. We let $T\subset \Sigma^*$  denotes a generic set of traces built on  alphabet $\Sigma$ and $t\in T\subset\Sigma^*$ a trace in $T$, for example, $t=\langle a,b,b,a,c\rangle\in T\subset\{a,b,c\}^*$.

\noindent
\textit{Stochastic language}. A stochastic language over an alphabet $\Sigma$ is a function $L:\Sigma^*\to[0,1]$ that provides the probability of the traces such that $\sum_{t\in\Sigma^*} L(t) = 1$. 

\noindent
\textit{Event log}. An event log $E$ is a multi-set of traces built on an alphabet $\Sigma$, i.e., $E\in Bag(\Sigma^*)$. Given a trace $t\in E$ we denote by $f(E,t)$ its multiplicity (i.e., its frequency in $E$). The stochastic language induced by an event log is straightforwardly obtained by computing the probability of the traces as $p(E,t)=f(E,t)/\sum_{t\in Supp(E)} f(E,t)$ where $Supp(E)$ denotes the set of unique traces in $E$, i.e., its support. The stochastic language of an event log $E$ will be denoted by $L_E$.

% For example  $L=\{\langle a,b,d,e\rangle ^{10}, \langle a,d,b,e\rangle ^{5}, \langle a,c,d,e\rangle ^2, \langle a,d,c,e\rangle\}$  is an event log  consisting of 10 traces $\langle a,b,d,e\rangle$, 5 traces $\langle a,d,b,e\rangle$, 2 traces $\langle a,c,d,e\rangle$ and one trace $\langle a,d,c,e\rangle$ for a total of 18 traces.

\noindent
\textit{Petri net}. A labelled Petri net model is a tuple $N=(P,T,F,\Sigma,\lambda,M_0)$, where $P$ is the set of places, $T$ the set of transitions, $F:(P\times T)\cup(T\times P)\to\mathbb{N}$ gives the arcs' multiplicity (0 meaning absence of arc), $\lambda: T\to (\Sigma\cup\tau)$ associates each transition with an action ($\tau$ being the silent action) and $M_0:P\to\mathbb{N}$ is the initial marking. For $N$ a PN We denote $RG(N)=(S,A)$, where $S=RS(N)$ is the reachability set (the set of  markings reachable from the initial one, and $A\subseteq RS(N)\times RS(N)\times T$  is the set of arcs  whose elements $(M,M',t)\in A$ are such that $M[t\rangle M'$, i.e. $M'$ is reached from $M$ by firing of $t$. 

%For $N$ a Petri net labelled w.r.t. an alphabet $\Sigma$  we denote $L_N\*=\{w\in \Sigma^* | \exists \sigma\in T^*, w=\lambda(\sigma)\}\subseteq \Sigma^*$ the language of $N$, where $\sigma$ is a sequence of firing transitions (i.e. corresponding to a path of $RG(N)$) and $\lambda(\sigma)$ is the sequence of labels of the transitions in $\sigma$.

\noindent
\textit{Workflow net}. A workflow net is a 1-safe\footnote{In any marking each place may contain at most 1 token } Petri net with the following structural constraints: 1) there exists a unique place, denoted \emph{source} with no incoming transition and a unique place denoted \emph{sink} with no outgoing transitions; 2) the initial marking is $M_0(source)=1$ and $M(p)=0$ for any place $p$ different from $source$ and 3) the net graph can be turned into a strongly connected  one by adding a single transition outgoing place $sink$ and ingoing place $source$.

\begin{figure}[htbp]
\begin{center}
\begin{tabular}{cc}
\begin{tabular}{c}
WN\\
\tikzstyle{place}=[circle,draw,minimum
size=4mm, rounded corners=0pt]

\begin{tikzpicture}
\node[place,tokens=1,label=above:{\footnotesize $source$}]        (source) {};
%\node[place,label=above:$p_2$,right=of source] (p2) {};

\node[node distance = 0.25 cm, transition, right=of source] (t1) {$a$};
\node[node distance = 0.5 cm, place,tokens=0, above right=of t1,label=above:$p_2$]        (p2) {};
\node[node distance = 0.5 cm,place,tokens=0, below right=of t1,label=above:$p_3$]        (p3) {};
\node[node distance = 0.25 cm, transition, right=of p2] (t2) {$b$};
\node[node distance = 0.25 cm,place,tokens=0, right=of t2,label=above:$p_4$]        (p4) {};    
\node[node distance = 0.5 cm, transition, below=of t2] (t4) {$c$};
\node[node distance = 0.5 cm, transition, below =of t4] (t5) {$d$};
\node[node distance = 1 cm,place,tokens=0, below=of p4,label=above:$p_5$]        (p5) {};    
\node[node distance = 2.25 cm, fill=black,transition, right =of t1] (t6) {$$};
\node[node distance = 0.25 cm,place,tokens=0, right=of t6,label=above:{\footnotesize $sink$}]        (sink) {};    
\draw (source) [->] to (t1);
\draw (t1) [->] to (p2);
\draw (t1) [->] to (p3);
\draw (p2) [->] to (t2);
\draw (t2) [->] to (p4);
\draw (p3) [->] to (t4);
\draw (p3) [->] to (t5);
\draw (t4) [->] to (p5);
\draw (t5) [->] to (p5);
\draw (p4) [->] to (t6);
\draw (p5) [->] to (t6);
\draw (t6) [->] to (sink);
\end{tikzpicture}\\
%{\footnotesize $L_1=\{\langle a,b,c\rangle^{15}, \langle a,c,b\rangle^{35}, \langle a,b,d\rangle^{15},\langle a,d,b\rangle^{35}  \}$}\\
%{\footnotesize $L_2=\{\langle a,b,c\rangle, \langle a,c,b\rangle, \langle a,b,d\rangle,\langle a,d,b\rangle,\langle a,d\rangle  \}$}
\end{tabular}
&
\begin{tabular}{c}
sWN\\
\tikzstyle{place}=[circle,draw,minimum
size=4mm, rounded corners=0pt]

\begin{tikzpicture}
\node[place,tokens=1,label=above:{\footnotesize $source$}]        (source) {};
%\node[place,label=above:$p_2$,right=of source] (p2) {};

\node[node distance = 0.25 cm, transition, right=of source,label=above:\scriptsize{$1$}] (t1) {$a$};
\node[node distance = 0.5 cm, place,tokens=0, above right=of t1,label=above:$p_2$]        (p2) {};
\node[node distance = 0.5 cm,place,tokens=0, below right=of t1,label=above:$p_3$]        (p3) {};
\node[node distance = 0.25 cm, transition, right=of p2,label=above:\scriptsize{$0.3$}] (t2) {$b$};
\node[node distance = 0.25 cm,place,tokens=0, right=of t2,label=above:$p_4$]        (p4) {};    
\node[node distance = 0.5 cm, transition, below=of t2,label=above:\scriptsize{$0.35$}] (t4) {$c$};
\node[node distance = 0.5 cm, transition, below =of t4,label=above:\scriptsize{$0.35$}] (t5) {$d$};
\node[node distance = 1 cm,place,tokens=0, below=of p4,label=above:$p_5$]        (p5) {};    
\node[node distance = 2.25 cm, fill=black,transition, right =of t1,label=above:\scriptsize{$1$}] (t6) {$$};
\node[node distance = 0.25 cm,place,tokens=0, right=of t6,label=above:{\footnotesize $sink$}]        (sink) {};    
\draw (source) [->] to (t1);
\draw (t1) [->] to (p2);
\draw (t1) [->] to (p3);
\draw (p2) [->] to (t2);
\draw (t2) [->] to (p4);
\draw (p3) [->] to (t4);
\draw (p3) [->] to (t5);
\draw (t4) [->] to (p5);
\draw (t5) [->] to (p5);
\draw (p4) [->] to (t6);
\draw (p5) [->] to (t6);
\draw (t6) [->] to (sink);
\end{tikzpicture}\\
%{\footnotesize $L_1=\{\langle a,b,c\rangle, \langle a,c,b\rangle, \langle a,b,d\rangle,\langle a,d,b\rangle  \}$}
\end{tabular}\\
\multicolumn{2}{c}{\footnotesize $L_1=\{\langle a,b,c\rangle^{15}, \langle a,c,b\rangle^{35}, \langle a,b,d\rangle^{15},\langle a,d,b\rangle^{35}  \}$}\\
\multicolumn{2}{c}{\footnotesize $M_1=\{\langle a,b,c\rangle^{0.15}, \langle a,c,b\rangle^{0.35}, \langle a,b,d\rangle^{0.15},\langle a,d,b\rangle^{0.35}  \}$}\\
\end{tabular}

\caption{An event log $L_1$ with corresponding workflow net (left) and  stochastic workflow net (right) whose transitions' weights are depicted above each transition: notice the  stochastic language  $M_1$ induced by the sWN conforms that of $L_1$.}
\label{fig:ex1}
\end{center}
\end{figure}
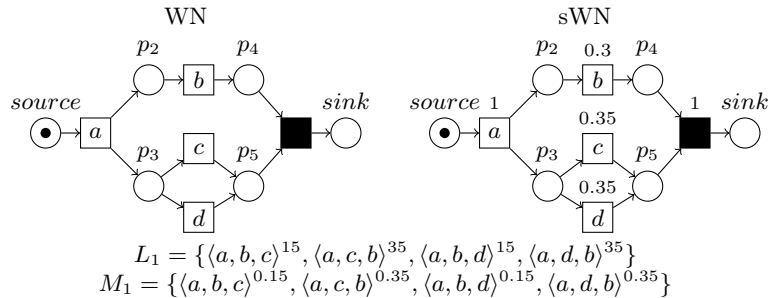

\noindent
\textit{Stochastic workflow net}. In the remainder we consider the stochastic, untimed,  extension of workflow nets, which we refer to as stochastic workflow nets (sWN). In practice a sWN $N_s=(P,T,F,W,\Pi,\Sigma,\lambda,M_0)$  is a Generalised Stochastic Petri Net (GSPN)~\cite{10.1145/288197.581193} consisting uniquely of \emph{immediate} transitions, i.e.  $T=T_i\cup T_t$, with $T_t=\emptyset$ ($T_t$, resp. $T_i$,   being the subset of timed transitions, resp. immediate transitions), each of which is associated with a non-negative weight $W:T\to \mathbb{R}_{>0}$ while priorities are all equal  ($\Pi(t)=1, \forall t\in T$), therefore in the remainder we omit $\Pi$ from the characterisation of   a sWN. Furthermore we use $RG(N_s)$ to refer to the reachability graph of a sWN $N_s$. We denote by $L_N$ the stochastic language associated with the sWN $N$.
%and $N_w=(P,T,F,\Sigma,\lambda,M_0)$ for the  WN embedded in $N_s$. 

\noindent
%\textit{Stochastic reachability  graph}. We consider an extended version of  reachability graph to deal with stochastic PN.   The stochastic RG of  a sWN $N_s=(P,T,F,W,\Sigma,\lambda,M_0)$ is denoted $SRG=(RG(N_w),\overline{P_T},\lambda)$, where $\overline{P_T}:RS(N_w)\times T\to[0,1]^{|T|}$ is the transitions weight vector. 

\begin{example}
Figure~\ref{fig:ex1} shows an example of event log, corresponding workflow net and stochastic workflow net together with the stochastic language induced by the latter. 
\end{example}

%\noindent
%\textit{Stochastic Earth Movers Conformance}.

\algrenewcommand\algorithmicrequire{\textbf{Input:}}
\algrenewcommand\algorithmicensure{\textbf{Output:}}

\section{Method}
\label{sec:method}
\begin{figure}
    \centering
   %%%%%%%%%%%%%%%%%%%%%%%% AUTOMATON-2 FOR REGION DISTANCE %%%%%%%%%%%%%%%%%%%
\begin{tikzpicture}[scale=.8,minimum width=1cm]
\renewcommand{\arraystretch}{.7}
   \everymath{\scriptstyle}
   
%% LOCATIONS
%\draw (-3,1.5) 
\node[left, initial text=,draw,rounded corners] (l0) { \begin{tabular}{@{}c@{}}\text{Event}\\  \text{Log} \end{tabular}};
%\draw (-1,1.5) 
\node[right=1cm of l0,draw,rounded corners] (l1) { \begin{tabular}{@{}c@{}}\text{Workflow}\\  \text{net} \\  \end{tabular} };
%\draw (1,1.5) 
\node[right=.5cm of l1,draw,rounded corners] (l2) { 
\begin{tabular}{@{}c@{}}\text{RG + }\\  \text{weights} \end{tabular}};

%\draw (3,1.5) 
\node[right=.5cm of l2,draw,rounded corners] (l3) { \begin{tabular}{@{}c@{}}\text{Stochastic} \\ \text{language}  \\  \text{via  unfolding}
\end{tabular} };

%\draw (5,1.5) 
\node[right=.5cm of l3,draw,rounded corners] (l4) { \begin{tabular}{@{}c@{}}\text{optimiser}\\ 
\end{tabular} };
\node[right=.5cm of l4] (l5){\begin{tabular}{@{}c@{}}\text{optimal}\\  \text{weights} \\\end{tabular}};

%% ARCS
\draw [-latex'] (l0) -- (l1) node [midway, above,sloped] {miner };
\draw [-latex'] (l1) -- (l2) node [midway, above,sloped] { };
\draw [-latex'] (l2) -- (l3) node [midway, above,sloped] { };
\draw [-latex'] (l3) -- (l4) node [midway, above,sloped] { };
\draw [-latex'] (l4) .. controls +(235:15mm) and +(315:15mm) .. (l3) node [pos=0.95 ,above] {};
\draw [-latex'] (l4) -- (l5) node [midway, above,sloped] {};
\draw [-latex',dashed] (l0) .. controls +(45:20mm) and +(130:20mm) .. (l4) node [pos=0.95 ,above] {};

{}; 
 
\end{tikzpicture}
    \caption{Optimised stochastic process discovery.}
    \label{fig:methodcartoon}
\end{figure}
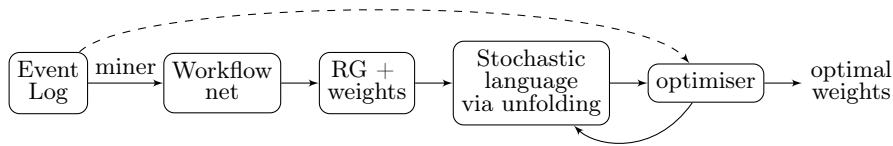

In the following we introduce a novel stochastic process discovery  procedure based on \emph{weight optimisation}. 
Figure~\ref{fig:methodcartoon} summarises  the relevant phases of the procedure. Starting from an event log, which contains the observations of the system under study, a WN model $N$ is obtained through some  mining algorithm and its corresponding RG is computed. 
The resulting RG, combined with an initial (random) vector of transition weights, is then used to evaluate the stochastic language issued by the stochastic interpretation of the WN, i.e. by associating the transitions of the mined WN with the weights of the considered weights vector.  
%and its stochastic extension is considered by adding a vector of weights for the transitions of the mined WN. % (e.g. inductive miner). 

%a method for computing the stochastic language associated with a sWN whose embedded WN has been  mined from an event. % log and assuming each transition $t_i$  of the Petri net is associated to a (real-valued) weight (symbolically) denoted $w_i$ (therefore we assume a stochastic interpretation of the mined Petri net). 

The assessment of the stochastic language that corresponds to a given vector of transitions' weights is achieved through breadth first unfolding of the RG (see Algorithm~\ref{alg:traceprob}). 
%The obtained stochastic language
 %The reachability graph   underlying the mined WN is then computed and from it the so-called \emph{stochastic language  graph} (SLG) is extracted through a   dedicated  breadth first unfolding  procedure. Each node of the obtained SLG corresponds to a finite prefix of the RG  (i.e. a finite word of the WN's language) and stores the probability of the corresponding prefix. 
 The obtained stochastic language is then fed in to an optimisation engine which by means of  a \emph{distance measure} assess the resemblance between the log's and model's stochastic languages and  iteratively search optimal weights for the transitions of the mined WN. %  by employing either a \emph{maximum-likelihood function} or a \emph{earth movers distance} as criteria to drive the optimisation search. 
 In the remainder we characterise the two main phases of the framework. 

\subsection{Computation of trace probability of a stochastic WN} %Construction of the stochastic language graph}

%\ah{it is not really a tree what we get} \pb{changed to graph}
%\ah{transitions have labels or actions?}
%\pb{transitions are labelled with actions of the event log}

The first step of the optimised stochastic discovery framework consists in computing the stochastic language associated with a given sWN. %  which is the task of Algorithm~\ref{alg:traceprob}. %We do so through a breadth first unfolding of the corresponding RG which in practice is implemented via a dedicated Queue. In order to deal with the specificity of traces unfolding in presence of \emph{silent transitions} the Queue's elements must  store dedicated relevant information (see description below).
It is done by \Call{TraceProbabilities}{$N,RG,U$} (shown in
Algorithm~\ref{alg:traceprob}) that takes as inputs a sWN $N$, its reachability graph $RG$ and a finite set of traces $U$, and returns  the probability that
traces of $U$ occur in $N$, in the form of set of pairs $(tr,pr)\in U\times [0,1]$, where $pr$ is the probability of trace $tr$. 

%by constructing a dedicated graph in whose nodes  the words belonging to the language issued by the sWN model are  associated with   their corresponding probability. 
%Before formalising the graph construction process (Algorithm~\ref{alg:slg}), we  define the graph like data structure it relies on. 

%\begin{definition}[Stochastic language graph]
%For $N_s=(P,T,F,W,\Sigma,\lambda,M_0)$ a sWN  and $T\subset \Sigma^*$ a finite set of traces built on alphabet $\Sigma$  we define the stochastic language graph of $N_s$ w.r.t. $L$ as $SLG(N_s,T)=(V,E)$, where $V\subseteq(S\times \mathbb{N} \times T)$ is the set of vertices with  $S=V(RG(N_w))$  the set of states of the reachability graph $(RG(N_w)$. A vertex $v\in V$ is a triple $v=(st,level,tr)$ where $st\in S$ is a state of the reachability graph $RG(N_w)$, $level$ represents the length of the unfolding path that lead to $st$ (from the initial state of $RG(N_w)$ and $tr\in prefixes(T)$ is a prefix of some trace in $L$ (i.e. $prefixes(T)$ being the set of all prefixes of traces in $L$).
%\end{definition}

%The SLG  is a data type  used in the procedure for  the computation of the probabilities of traces belonging to the stochastic language of the considered sWN. Such computation is achieved by breadth first unfolding of the corresponding RG which is obtained via a dedicated Queue whose items are pairs $(v,pr)\in V\times [0,1]$

\begin{algorithm}
\caption{Reachability graph unfolding to compute trace probabilities\label{alg:unfold}}
\begin{algorithmic}[1]
\Statex $\!\!\!\!\!\!\!\!\!\!$\Call{TraceProbabilities}{$N,RG,U$}
\Require a  sWN $N=(P,T,F,W,\Sigma,\lambda,M_0)$, the corresponding reachability graph $RG=(S,A)$, a finite set of traces $U\in\Sigma^*$
\Ensure $D\subseteq U\times [0,1]$ set of pairs associating traces of $U$ with probability value
  \State $\mathit{PrArcs}\leftarrow \{\} $
  \For {all $(st,st',t)\in A$} \label{alg:prarcinit}
  \State $sumw=0$ % $prarc\leftarrow W(t)$, $sumw=1$
  \For {all $(st,st'',t')\in A$}
  \State $sumw \leftarrow sumw + W(t')$ %$prarc\leftarrow prarc \cdot sumw/(sumw+w(t)) $, $sumw \leftarrow sumw+w(t')$
  \EndFor \label{alg:prarcend}
  \State $\mathit{PrArcs} \leftarrow \mathit{PrArcs}\cup \{(st,st',t,W(t)/sumw)\}$
  \EndFor
  \State $D\leftarrow \{\}$, $Q \leftarrow$ empty queue
  %\State $Q \leftarrow$ empty queue
  \State $Q$.\Call{Enqueue}{$~((\mbox{initial~state~of}~RG, 0, \epsilon),1)~$}   \label{alg:initenque}
  \While{$Q$ is not empty}
  \State $((st, level, tr),pr)~\leftarrow$ $Q$.\Call{Dequeue}{$~\!$} \label{alg:dequeue}
  \For{all   $(st,st',t,prarc)\in \mathit{PrArcs}$} \label{alg:outg}
  %\State $newst~\leftarrow~a.tostate$ \label{alg:newst}
  \State $newst~\leftarrow~st'$ \label{alg:newst}
  \State $newtr~\leftarrow~ tr+\lambda(t)$ 
  \State $newpr~\leftarrow pr \cdot prarc$  \label{alg:newprob}
  \If{$newst$ is the sink} \label{alg:sink}
  \If{$newtr$ is a trace in $U$}
  \State $D$.\Call{AddOrIncrease}{$newtr$, $newpr$}
  \EndIf \label{alg:endsink}
  \ElsIf{$newtr \in prefixes(U)$}
  \State $Q$.\Call{AddOrIncrease}{$(newst,level+1,newtr)$, $newpr$}
  \EndIf
  \EndFor
  \EndWhile
  \State\Return{$D$}
\end{algorithmic}
\label{alg:traceprob}
\end{algorithm}

The algorithm begins  (line~\ref{alg:prarcinit} to~\ref{alg:prarcend}) with enriching each arc of the RG with the probability of the corresponding  transition computed w.r.t. the   weight function of the sWN. 
 The probability of each arc of the RG is needed, in the remainder (line~\ref{alg:newprob}), for  establishing  the probability of the  paths as they are unfolded.
 
 In order to follow a breadth first order for unfolding  $RG(N_s)$, a queue, $Q$, is used. 
  The elements of the queue are pairs $((st,level,tr), pr )$ 
whose first element, is a triple $(st,level,tr)$, where $st\in S$ is a state of the $RG(N_s)$, $level\in\mathbb{N}$ represents the length of the unfolded path that leads to $st$ (from the initial state of $RG(N_s)$) and $tr\in prefixes(U)$ is a prefix of some trace in $U$ ($prefixes(U)$ being the set of all prefixes of traces in $U$). The second element $pr\in [0,1]$ is the probability of the path.
%in which a node of the SLG (i.e. a triple $s,n,t$) is  associated with a probability value (i.e. the probability of trace $t$ to occur in the model $N_s$). The triple
%contains the {\em state} that was reached by the so far unfolded arcs (transitions) of the RG, the {\em number of arcs} to reach the state and the {\em trace} produced along the arcs. 
$Q$ is initialised (line~\ref{alg:initenque}) by enqueuing 
 $((source, 0, \epsilon),1)$ meaning the initial triple,  corresponding to the initial state $source$ of the RG, length 0 and with the empty trace,  is assigned  probability one.

 The probability of the traces in $U$ are kept in a set of
pairs, $D$, composed of a trace and a real number. For both the queue, $Q$,
and the set, $D$, we assume to have a procedure we call
\Call{AddOrIncrease}{key $k$, value $v$} that adds $k$ with value $v$ if
$k$ is not present, otherwise it increases the value associated with $k$ by
$v$. In case of $Q$ the key is a triple $(state, level, trace)$ while in
case of $D$ is a $trace$. This mechanism is necessary because in most cases,
due to the silent transitions, the same triple in $Q$ (trace in $D$) can be
reached (generated) through different sequences of transitions.

The unfolding procedure, shown in Algorithm~\ref{alg:unfold}, proceeds by
dequeuing a triple from $Q$ (line~\ref{alg:dequeue}) and exploring all
outgoing arcs of the corresponding state of the RG (line~\ref{alg:outg}).
From line~\ref{alg:newst} to \ref{alg:newprob} we take into account the
effect of the considered arc.  From line~\ref{alg:sink} to
\ref{alg:endsink}, we cover the case in which the newly reached state is
the $sink$: if the generated trace is present in $U$ then the pair
$(trace, probability)$ is either added to $D$ or the associated value is
updated. If the newly reached state is not the $sink$ and the trace
generated so far is a prefix of a trace in $U$, we either add a new item to
the queue or increase the probability associated with an existing one. When
the queue is empty $D$ is returned.

It is worth to mention that in order to proceed as described above, it is
necessary to include in the items of the queue the number of traversed
arcs.  This is due to the fact the mined WNs with their silent transitions
are usually such that the same $(state, trace)$ pair can be reached by
different number of arcs. Keeping track of the number of traversed arcs
helps to proceed in breadth first fashion and not to loose information on
the way

Algorithm~\ref{alg:unfold} is presented in such a way that the unfolding is
restricted to a set of traces, which is useful to evaluate traces present
in an event log. This is an advantageous choice from the point of view of
execution time since event logs usually contain much less traces than the
number of traces their mined WN counterpart can produce. Moreover, most
mined WNs contain cycles leading to an infinite number of possible traces.
It is however straightforward to modify the algorithm to make the
calculations not based on a set of traces. If the number of possible traces
is infinite, one can put a limit on the length of the traces (disregarding 
the silent transitions) or cover a given amount of probability, for
example, so to stop the unfolding when the total probability in $D$ is more than
0.9. It is also possible that, due to the complexity of the WN or the
characteristics of the event log, even restricting the calculation to the
traces present in the event log the execution time is unfeasible (this can
easily happen in the presence of long traces that can be generated by many
different transition sequences of the WN). In this case a limit on the
length of the traces can be introduced to obtain at least an
approximation.

%%%%%%%%%%%%%%%%%%%%%%%%%%%%%%
\begin{example}
We illustrate the function of the computation of the stochastic language of a sWN that corresponds to a WN mined from a log by application of Algorithm~\ref{alg:traceprob}. 
Figure~\ref{fig:exampleminedPN}
shows a WN   mined~\footnote{using the inductive miner algorithm~\cite{DBLP:conf/apn/LeemansFA13}.} from the event log
 $$E=
\{\langle AAAA\rangle, \langle AAA\rangle, \langle QAQAQ\rangle, \langle AA\rangle, \langle AAQQA\rangle\}$$ built on alphabet $\Sigma=\{A,Q\}$. The WN contains seven silent transitions and two
labelled with the activities in $\Sigma$.
The corresponding reachability graph is
shown in Figure~\ref{fig:exampleminedRG} (states are labelled with the corresponding WN marking depicted as the string composed by appending the  names of places which contain a token). Note that the language associated
with the WN is not finite since the WN contains loops involving non-silent transitions. 

\begin{figure}[h!]
\centerline{\includegraphics[width=\columnwidth]{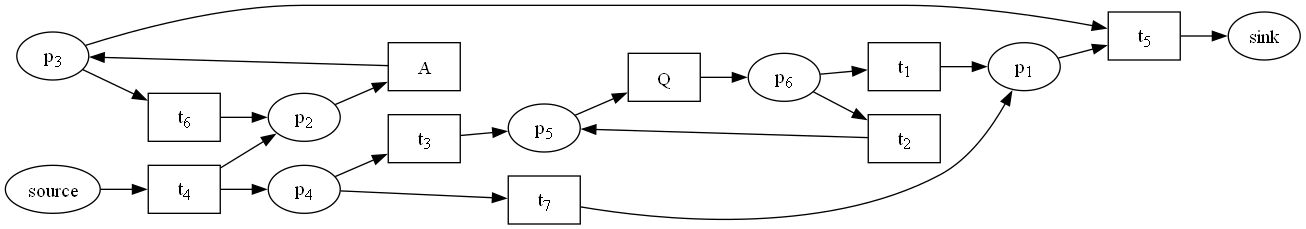}}
\caption{WN mined from log $E=
\{\langle AAAA\rangle, \langle AAA\rangle, \langle QAQAQ\rangle, \langle AA\rangle, \langle AAQQA\rangle\}$} \label{fig:exampleminedPN}
\end{figure}

\begin{figure}[h!]
\centerline{\includegraphics[width=0.8\columnwidth]{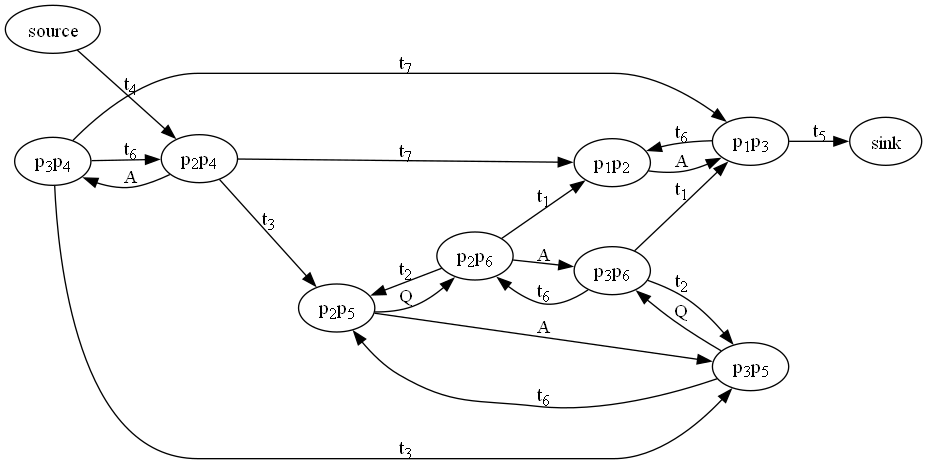}}
\caption{Reachability graph of the WN in Figure~\ref{fig:exampleminedPN}.\label{fig:exampleminedRG}}
\end{figure}

Figure~\ref{fig:exampleunfolding} depicts the directed acyclic graph resulting from the unfolding of the RG  by application of Algorithm~\ref{alg:traceprob}. Every node is labelled with the
corresponding marking and the trace corresponding to the non-silent transitions occurred in the path  from the
initial state to the considered node (notice that for simplicity, in Figure~\ref{fig:exampleunfolding} , the length of the path is not depicted for nodes as nodes are  positioned level by level, with the initial node being at level 0). Note that, for sake of completeness, Figure~\ref{fig:exampleunfolding} contains  a few
nodes that, during the execution of Algorithm~\ref{alg:traceprob}, are not inserted neither in the queue $Q$, nor in the set
$D$, as their corresponding trace does not belong to the considered event log.  For example the left most node
in level 6, i.e.,  ($sink$,$QA$), resulting from a transition sequence that arrives
to the $sink$ producing the trace  $\langle QA\rangle\not\in
\{\langle AAAA\rangle, \langle AAA\rangle, \langle QAQAQ\rangle, \langle AA\rangle, \langle AAQQA\rangle\}$. 

\begin{figure}[h!]
\centerline{\includegraphics[width=\columnwidth]{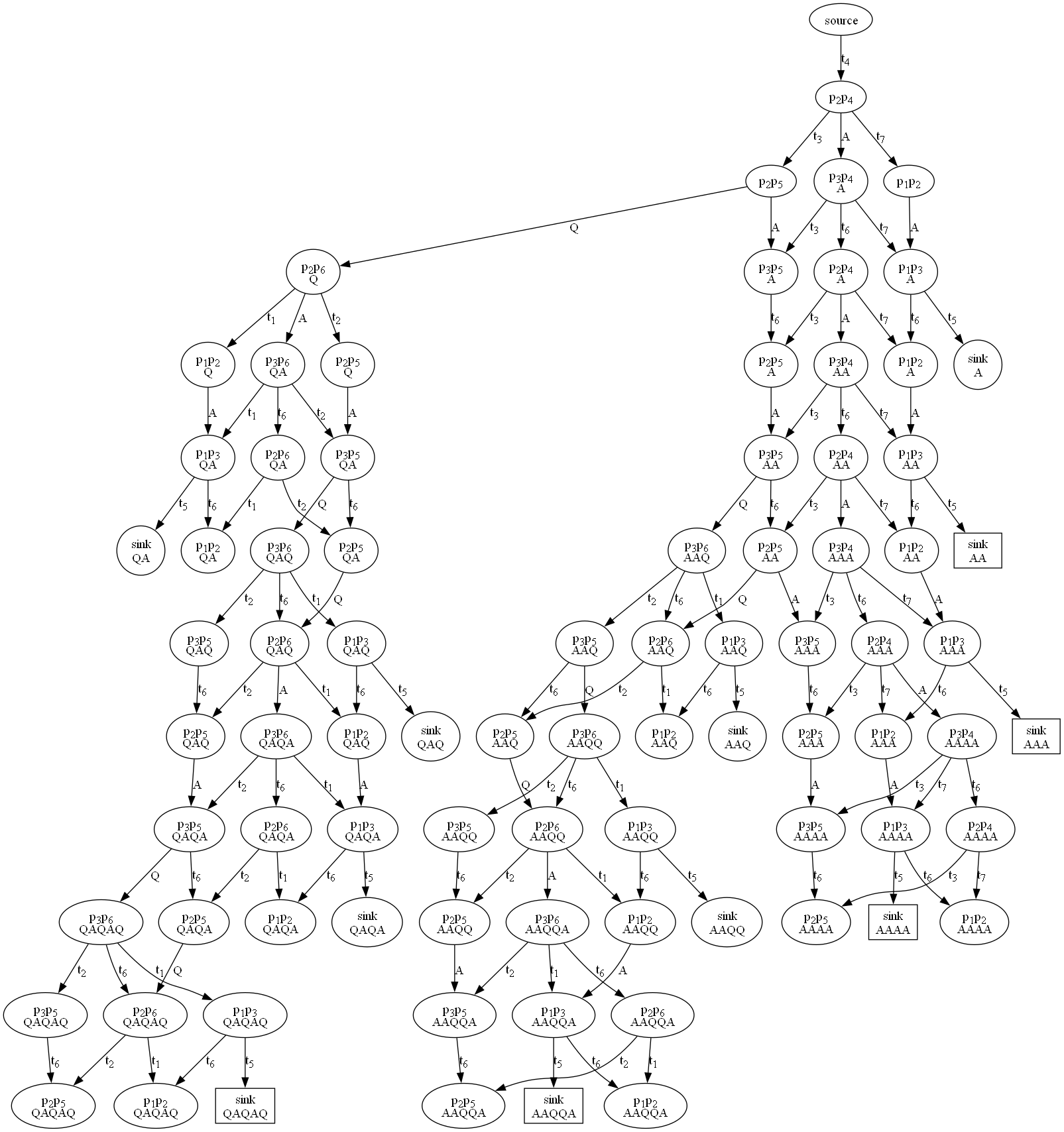}}
\caption{Unfolding of the RG in Figure~\ref{fig:exampleminedRG} restricted
  to the traces $AAAA$, $AAA$, $QAQAQ$, $AA$ and $AAQQA$. Rectangles
  correspond to completed traces present in the
  event log.\label{fig:exampleunfolding}}
\end{figure}

In order to give a flavor of the expressions constructed by the unfolding,
we report here the equations needed to calculate the probability of two
traces, one present in the event log, $AA$, and the other not, $QA$.  The
probability of the nodes that are necessary to calculate $P(AA)$ and
$P(QA)$ are listed level by level from left to right (see
Figure~\ref{fig:exampleunfolding}):
{\scriptsize
\begin{align*}&
P(source, 0) = 1,~~
P(p_2p_4, 1) = P(source, 0),~~
P(p_2p_5, 2) = P(p_2p_4, 1) \frac{w_3}{w_3 + w_7 + w_A},~~ \\ &
P(p_3p_4, 2, A) = P(p_2p_4, 1) \frac{w_A}{w_3 + w_7 + w_A},~~
P(p_1p_2, 2) = P(p_2p_4, 1) \frac{w_7}{w_3 + w_7 + w_A},~~\\ &
P(p_2p_6, 3, Q) = P(p_2p_5, 2) \frac{w_Q}{w_A + w_Q},~~
P(p_2p_4, 3, A) = P(p_3p_4, 2, A) \frac{w_6}{w_3 + w_6 + w_7},~~\\&
P(p_1p_3, 3, A) = P(p_3p_4, 2, A) \frac{w_7}{w_3 + w_6 + w_7} + P(p_1p_2, 2),~~\\&
P(p_1p_2, 4, Q) = P(p_2p_6, 3, Q) \frac{w_1}{w_1 + w_2 + w_A},~~
P(p_3p_6, 4, QA) = P(p_2p_6, 3, Q) \frac{w_A}{w_1 + w_2 + w_A},~~\\&
P(p_3p_4, 4, AA) = P(p_2p_4, 3, A) \frac{w_A}{w_3 + w_7 + w_A},~~\\&
P(p_1p_2, 4, A) = 
P(p_2p_4, 3, A) \frac{w_7}{w_3 + w_7 + w_A} +
P(p_1p_3, 3, A) \frac{w_6}{w_5 + w_6},~~\\&
P(p_1p_3, 5, QA) = 
  P(p_1p_2, 4, Q) + P(p_3p_6, 4, QA) \frac{w_1}{w_1 + w_2 + w_6},~~\\&
P(p_1p_3, 5, AA) = 
  P(p_3p_4, 4, AA) \frac{w_7}{w_3 + w_6 + w_7} + P(p_1p_2, 4, A),~~\\&
P(sink, 6, QA) = P(p_1p_3, 5, QA) \frac{w_5}{w_5 + w_6},~~
P(sink, 6, AA) = P(p_1p_3, 5, AA) \frac{w_5}{w_5 + w_6}
\end{align*}}
from which
{\scriptsize
\begin{align*}
  P(QA)&=
  \frac{w_1 w_3 w_5 (w_1 + w_2 + w_6 + w_A) w_Q}
       {(w_1 + w_2 + w_6) (w_5 + w_6) (w_1 + w_2 + w_A) (w_3 + w_7 + w_A) (w_A + w_Q)}\\
  P(AA)&=
  \frac{w_5 w_6 w_7 (w_3 + w_6 + w_7 + 
   w_A) ((w_3 + w_7) (w_3 + w_6 + w_7) + (w_3 + w_5 + 2 w_6 + w_7) w_A)}
  {(w_5 + w_6)^2 (w_3 + w_6 + w_7)^2 (w_3 + w_7 + w_A)^2}
\end{align*}}
\noindent 
which we show because it outlines the complicated role that each parameter
can have in the probabilities involved in the parameter estimation.

%We did not detail in Algorithm~\ref{alg:unfold} the calculation of the probability that a given outgoing arc of a state is chosen in the RG. Such probability is simply the weight of the transition corresponding to the arc divided by the sum of the weights of the transitions corresponding to all outgoing arcs. From a computational perspective it is convenient to calculate the probability of every arc of the RG in advance.
\end{example}
%%%%%%%%%%%%%%%%%%%%%%%%%%%%%%

\subsection{Measuring the distance between two stochastic languages}

The main goal of this paper is {\em weight estimation}, i.e., to find such a weight function, $W$, for the mined sWN, $N$, with which the stochastic language produced by the sWN, $L_N$, resembles that of the
event log, $L_E$. This requires to choose a way to measure the distance (called also divergence or difference) between two stochastic
languages, namely, $L_N$ and $L_E$. In order to make explicit the role of the weight function, the probability that $N$ produces a given trace $t$ will be denoted by $L_N(t,W)$. 

The use of one distance measure, namely, the {\em Kullback-Leibler divergence} (KLD) (called also relative entropy), corresponds to the probably most used parameter estimation approach which is {\em maximum likelihood estimation}. The maximum likelihood estimate corresponds to that point in the parameter space, in our case those transition weights, with which the observed data is most likely. The corresponding optimization procedure is based on the likelihood function which, given the transition weights, determines how probably the sWN $N$ reproduces $L_E$.

The likelihood function, whose maximization corresponds to the minimization of the KLD between $L_E$ and $L_N$, is
\begin{equation}
  \label{eq:ll}
  M_{LH}(W)=\prod_{t\in Supp(E)} L_N(t,W)^{L_E(t)}
\end{equation}
Since \eqref{eq:ll} can easily be subject to numerical underflow, to optimization purposes usually its logarithm is used (which clearly shares the position of the maximum) 
\begin{equation}
  \label{eq:logllopt}
  \log(M_{LH}(W))=
  \sum_{t\in Supp(E)} L_E(t) \log(L_N(t,W))
\end{equation}
It is important to note that the calculation of \eqref{eq:logllopt} requires the probability of only those traces that are present in the event log $E$.

Another way to measure the difference between two distributions is
provided by the popular {\em earth moving distance} (EMD), called also {\em Wasserstein distance} (WD). Intuitively, the idea is to compute the minimal ``cost'' to transform one distribution into the other,  where cost is intended as the {\em amount of probability} that has to be moved multiplied by the average {\em distance} it needs to be moved.

\begin{figure}
\begin{minipage}{0.49\textwidth}
  \includegraphics[width=\columnwidth]{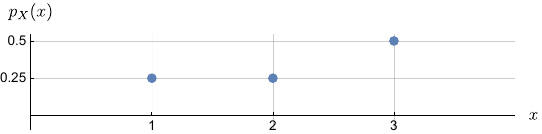}
\end{minipage}
\hfill
\begin{minipage}{0.49\textwidth}
  \includegraphics[width=\columnwidth]{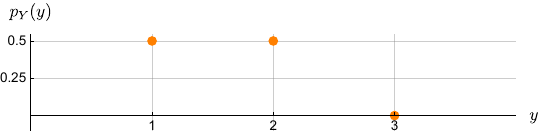}
\end{minipage}
  \caption{Probability mass functions $p_X(x)$ and $p_Y(y)$.\label{fig:pmf1}}
\end{figure}

As an example, consider two discrete random variables, $X$ and $Y$, with support 
$\{1,2,3\}$ and probability mass functions, $p_X(x)$ and $p_Y(y)$, as
depicted in Figure~\ref{fig:pmf1}.
In this simple case, it is obvious that, in order to transform $p_X$ to
$p_Y$ with minimal cost, 1/4 of probability needs to be moved from $x=3$ to
$y=1$ with associated distance $3-1=2$ and 1/4 of probability from $x=3$ to
$y=2$ with associated distance $3-2=1$. The total cost is $1/4 \cdot 2+1/4
\cdot 1=3/4$ which is the EMD distance between $p_X$ and
$p_Y$. Transforming $p_Y$ to $p_X$ has the same minimal cost.  The
transformation can be described by the so-called transport matrix ($M$) and the
associated distance matrix ($C$) which are, respectively,
\begin{equation}
\label{eq:tmatrix}
M=\begin{pmatrix}
1/4 &0 &0 \\
0 &1/4&0\\
1/4 & 1/4 &0
\end{pmatrix}
~~C=\begin{pmatrix}
0 &1 &2 \\
1 &0&1\\
2 & 1 &0
\end{pmatrix}
\end{equation}
where reading $M$ row-wise give the transformation from $p_X$ to $p_Y$
while column-wise from $p_Y$ to $p_X$. Note also that row sums provide
$p_X$ while column sums $p_Y$.

In case of  distributions  
defined over a metric space,  such as the above given $p_X$ and $p_Y$,
the distance
associated with moving probability from a point $x$ to a point $y$ is
naturally defined, i.e., it is $|x-y|$. However, in the case of  stochastic
languages, a notion of  distance between two traces (words)
must be considered. The most natural and practically  applied one 
 is the Levenshtein distance which measures the distance between two
strings (traces composed of actions in our case) as the minimum \emph{alignment}~\cite{https://doi.org/10.1002/widm.1045}, that is, the minimum number of
single-character edits (insertion, deletion or substitution of an action)
needed to change one word (trace) into the other. The Levenshtein distance
between two traces, $t_1$ and $t_2$, will be denoted by $c(t_1,t_2)$.

In order to define the EMD in our context, we denote by $T_N$ the set of traces generated by the sWN $N$ (the fact that it can be infinite will be dealt with later). 

Then the EMD between $L_E$ and $L_N$ is defined based on the following optimization problem (recall that we must find the way that transforms one into the other with minimal
cost):
\begin{align}
  \label{eq:EMDistance}
&  M_{EMD}(W)=\min_m \sum_{t_1\in Supp(E),t_2 \in T_N} c(t_1,t_2)m(t_1,t_2)\\
\nonumber
  &  \mbox{such that}~
  \left\{\begin{array}{lr}
  \sum_{t_2 \in T_N} m(t_1,t_2)=L_E(t_1)&\forall t_1 \in Supp(E) \\
  \sum_{t_1 \in Supp(E)} m(t_1,t_2)=L_N(t_2,W)&\forall t_2 \in T_N \\
  m(t_1,t_2)\geq 0&\forall t_1 \in T_1, t_2 \in T_2
  \end{array}\right.
\end{align}
where the function $m$ is analogous to the transport matrix in
\eqref{eq:tmatrix}. Using a normalized version of the Levenshtein distance,
it is guaranteed that $0 \leq M_{EMD}(W) \leq 1$,  resulting in a measure
that, conversely to both likelihood function and Kullback-Leibler divergence, is possible to interpret (i.e. 0 indicates   perfect
matching while 1  complete difference). 
If $Supp(E)$ and $T_N$ are finite sets computing \eqref{eq:EMDistance}
corresponds to a linear programming problem whose size depends on $|Supp(E) \cup T_N|$.

\noindent
\emph{Dealing with infinite languages}. 
In case the net language $T_N$ is infinite, which is common as mining algorithms often yields nets containing cycles, the exact evaluation of the EMD is replaced by the so called \emph{truncated} EMD (tEMD), commonly expressed as tESMC, i.e., its 1-complement, tEMSC=1-tEMD~\cite{10.1007/978-3-030-26643-1_8}. The basic idea behind tEMD computation is to consider a sufficiently large subset of traces that covers a given amount of
probability. It turns out that even relatively small ``coverage'' thresholds, e.g.,
0.8, may correspond to a large number of traces to be unfolded hence  making  tEMD calculation slow (see experiments in Table~\ref{tab:exp1})  and therefore often not suitable for the purpose of parameter estimation.
%Since $T_N$ is usually infinite due to cycles in the sWN $N$, we consider a finite subset to calculate the EMD. One approach, used, for example, in \cite{10.1007/978-3-030-26643-1_8}, is to use a subset that covers a given amount of probability. It turns out that even relatively small ``coverage'', like 0.8, can correspond to a large number of traces which makes the calculation of the EMD slow. Consequently, this way of calculating the EMD, which we will refer to as truncated earth moving ditance (tEMD), is not feasible for the purpose of parameter estimation.

As an alternative, in this paper we propose a different approach, which we
name \emph{restricted} EMD (rEMD), that is: to restrict the calculation of
the EMD to those traces that are present in the event log. This, in most
cases, results in a defective stochastic language, i.e., the total
probability is less than one. In order to calculate the EMD
\eqref{eq:EMDistance}, we apply normalization to obtain a non-defective
stochastic language. In most cases, $|Supp(E)|\ll |T_N|$ even when $T_N$ is
not infinite. Hence reducing the calculation of the EMD to the traces in
$Supp(E)$ can be feasible to optimization purposes.

Finally we point out that 
the  distance measures $\log(M_{LH}(W))$, given in \eqref{eq:logllopt},  and $M_{EMD}(W)$, given in \eqref{eq:EMDistance}, can be calculated based on the probabilities returned by \Call{TraceProbabilities}{} given in Algorithm~\ref{alg:unfold}.

\begin{algorithm}[]
\caption{Weight optimization\label{alg:optw}}
\begin{algorithmic}[1]
\Statex $\!\!\!\!\!\!\!\!\!\!$\Call{OptimizedWeights}{$m,n_0,max_{iter}$}
\Require distance measure $m$ (LH or EMD) and two integers $n_0>0$ and $max_{iter}>0$
\Ensure the optimized weights $W_{opt}$
  \For{$i=1,2,...,n_0$}
  \State $W \leftarrow$ random weights  
  \If{$m=$ LH}
      \State $M \leftarrow -\log(M_{LH}(W))$
  \Else
      \State $M \leftarrow M_{EMD}(W)$
\EndIf
  \If{$M<M_{best}$ or $i=1$}
  \State $M_{best} \leftarrow M$, $W_{0}\leftarrow W$
\EndIf
\EndFor
  \If{$m=$ LH}
      \State $W_{opt} \leftarrow$ \Call{Minimize}{$-\log(M_{LH}(W)),W_{0},max_{iter}$} 
  \Else
      \State $W_{opt} \leftarrow$ \Call{Minimize}{$M_{EMD}(W),W_{0},max_{iter}$} 
  \EndIf
  \State\Return{$W_{opt}$}
\end{algorithmic}
\label{alg:weightopt}
\end{algorithm}

\subsection{Weight optimization}

Optimization of the weight function $W$ (which, in an implementation, is represented simply as a vector of positive real-valued weights) can be performed either by maximizing $\log(M_{LH}(W))$ \eqref{eq:logllopt} or by minimizing $M_{EMD}(W)$ \eqref{eq:EMDistance}. In order to minimize in both cases one can simply use $-\log(M_{LH}(W))$ instead of $\log(M_{LH}(W))$. We experimented with several constrained gradient descent algorithms that requires a starting point in the parameter space. Here we refer to these algorithms in general as \Call{Minimize}{$f_{obj}(W),W_{0},max_{iter}$} where $f_{obj}(W)$ is the objective function, $W_0$ is the starting point and $max_{iter}$ is the maximal number of iterations (the optimization might stop before if it is not able to further minimize $f_{obj}(W)$)).
\Call{Minimize}{} returns the optimized weights.
In Section~\ref{sec:results} we give more details on the applied optimization algorithms.
The objective function is such that several local optimums may exist in which the optimization process can get stuck. For this reason, and also to speed up optimization, we begin with choosing the most promising starting point among $n_0$ randomly selected points of the parameter space. The corresponding algorithm is formulated in such a way that the choice between LH and EMD depends on a parameter.
It is shown in Algorithm~\ref{alg:optw}. 

\subsection{Execution time}

Precise analysis of the computational complexity of the applied algorithms is highly difficult due to the large number of factors it depends on. We try to list here the main factors and their effect.

During the unfolding (Algorithm~\ref{alg:unfold}) the queue $Q$ contains elements $ (state$, $level$, $trace)$ corresponding to the same $level$ or elements corresponding to two consecutive levels. The potential number of elements with a given $level$ depends on the number of markings that can be reached in $level$ transition firings (which is related to the number of markings of the sWN) and on the number of different traces that can be generated arriving to a given marking. The way this number grows can heavily depend on the size of the alphabet. Larger alphabet usually means slower unfolding. The structure of the mined sWN is also crutial, the larger the number of concurrently enabled transitions the slower the unfolding. Often the mined sWN is such that firing of a set of transitions in any order starting from a marking leads to the same marking. Also this has a negative effect on the execution time. On the other hand, limiting the unfolding to traces of an event log usually makes the unfolding feasible as it restricts it  to a significantly lower number of branches.  

Once the probability of the traces is obtained by the unfolding, the calculation of $\log(M_{LH}(W))$ is immediate even for large event logs since it corresponds to a simple sum. Instead, in case of using $M_{EMD}(W)$ the size of the event log is decisive and can be prohibitive since calculating $M_{EMD}(W)$ itself is an optimization problem.

For what concerns the optimization of the weight, the experimented gradient descent algorithms have an execution time that is proportional to the time needed to calculate the objective function (this requires unfolding and calculation of $\log(M_{LH}(W))$ or $M_{EMD}(W)$), to the number of transitions (that is the number of weights to optimize) and to the number of iterations (which we limit by the parameter $max_{iter}$).

Execution times of the different phases of the overall optimization procedure are given in Section~\ref{sec:results}.

\section{Results}
\label{sec:results}
We have realised  a prototype software tool~\footnote{The source code, datasets used for experimentation, and the obtained results are publicly available at the Git repository at \gitlink. 
%The experiments described here were conducted on a PC running Linux Ubuntu 20.04.6 LTS, equipped with a 2.60GHz CPU and 32 GB of memory and using Python 3.9 interpreter. 
}, written in Python,   that implements the weight optimisation procedure outlined in  Algorithm~\ref{alg:weightopt} via  \Call{TraceProbabilities}{} computation given in Algorithm~\ref{alg:traceprob}. 
In our implementation of the optimisation procedure we have added a further   ``convergence''  parameter $\delta$ to control the ending of   parameters  search. 
%Therefore  one can either opt between letting the minimisation phase run up to $max_i$ iterations (i.e. calling the $minimize$ function of Algorithm~\ref{alg:weightopt}  with $max_i>1$), or to control the effect of each (single) minimisation step 
Therefore the search for optimal weights stops either after $max_{iter}$ iterations of the minimisation phase (i.e. the \Call{Minimize}{} function of Algorithm~\ref{alg:weightopt}) or as soon as  the minimised distance $m$ (either $LH$ or $EMD$) has converged within  the $\delta$ threshold (i.e., at each iteration we check that $|m_i-m_{i-1}|/m_i < \delta | $, where $m_i$ is the  value of distance $m$ at iteration $i$). 
%Specifically, with such modality, we allow for a single iteration of the minimisation phase (i.e. the $minimize$ function of Algorithm~\ref{alg:weightopt} is called with $max_i=1$) and we determine whether the minimised distance $m$ (either $LH$ or $EMD$) has converged within  the $\delta$ threshold (i.e. we check that $|m_i-m_{i+1}|/m_i < \delta | $, where $m_i$ is the  value of distance $m$ at iteration $i$). 

In order to test it   we have considered different real-life event logs, most  of which were made available through the Business Process Intelligence (BPI) challenge and are freely accessible at \url{https://data.4tu.nl/.} The logs have different complexity 
 and yield WN models and corresponding RGs with different complexity (see Table~\ref{tab:timeprop}). In all of our experiments (except those corresponding to column 'RSD') we have used the inductive miner algorithm~\cite{DBLP:conf/apn/LeemansFA13} to mine the WN from each log and then  estimated the transition weights to obtain a corresponding sWN.

In Table~\ref{tab:exp1} we compare log-likelihood (LH), restricted EMD (rEMD) and  truncated EMD (tEMD) distances obtained through our optimisation scheme  with those obtained through Burke's six estimators~\cite{DBLP:conf/icpm/BurkeLW20} and also with that obtained by applying the Rogge-Solti stochastic process discovery approach~\cite{5f0e8dd04572478fb450eefe4211d69f}. 

%%%%%%%% TABLE 1 %%%%%%%%%%
\begin{table}[ht]
    \centering
    \resizebox{\textwidth}{!}{%
    \begin{tabular}{l p{1ex} l l l p{1ex} l l l p{1ex} l l l l p{1ex} l}
        \toprule 
        Event log & & \multicolumn{3}{l}{LH-opt} & & \multicolumn{3}{l}{rEMD-opt} & & \multicolumn{4}{l}{Best estimator} & & RSD \\
        \cmidrule(){3-5} \cmidrule(){7-9} \cmidrule(){11-14} \cmidrule(){16-16}
         & & LH & rEMD & tEMD & & LH & rEMD & tEMD & & name & LH & rEMD & tEMD & & tEMD \\
        \hline
        BPIC13\_c & & 4.05 & 0.14 & 0.12 & & 5.17 & 0.04 & TO & & fork & 9.85 & 0.63 & TO & & 0.67 \\
        BPIC13\_i & & 9.15 & 0.32 & TO & & 10.72 & 0.18 & TO & & lhpair & 15.22 & 0.70 & TO & & 0.86 \\
        BPIC13\_o & & 3.81 & 0.10 & 0.07 & & 4.49 & 0.076 & 0.11 & & pairs & 5.22 & 0.24 & TO & & 0.51 \\
        BPIC17\_ol & & 1.98 & 0.06 & 0.05 & & 2.53 & 0.09 & 0.13 & & freq & 5.81 & 0.09 & 0.1 & & 0.08 \\
        BPIC20\_dd & & 3.55 & 0.10 & 0.13 & & 9.04 & 0.02 & TO & & fork & 25.14 & 0.93 & TO & & TO \\
        BPIC20\_rfp & & 4.85 & 0.15 & TO & & 34.92 & 0.41 & TO & & freq & 97.09 & 0.99 & TO & & TO \\
        Roadfines & & 3.82 & 0.14 & TO & & 8.23 & 0.08 & TO & & pairs & 5.99 & 0.27 & TO & & TO \\
        \bottomrule
    \end{tabular}
    }
    \caption{Comparing optimised/non-optimised weight estimation on real-life  logs.}
    \label{tab:exp1}
\end{table}

%%%%%%%%%%%%%%%%%%%%%%%%%%%

%results obtained  For each event log we have computed our optimisation procedure using both maximum-likelihood as well as EMD (limited to the log's traces)

The 'LH-opt' columns refer to results obtained by optimisation  w.r.t the LH measure~(\ref{eq:ll}) while the 'rEMD-opt' columns  to optimisation of the EMD distance~(\ref{eq:EMDistance}) using the restricted EMD measure. In practice for LH optimisation we resorted to   the 'L-BFGS-B' method~\cite{10.1145/279232.279236}  while for EMD optimisation to  the 'Powell' method~\cite{10.1093/comjnl/7.2.155} both available  through the \texttt{minimize} function  of Python's \texttt{scipy.optimize} class. These two methods provided the best performance among the many provided by \texttt{minimize} for LH optimization and EMD optimization, respectively.

%For EMD minimisation we opted for the Powell's method as it allows for a faster convergence (w.r.t. L-BFGS-B).  

Results in the 'LH-opt' and 'rEMD-opt' columns of Table~\ref{tab:exp1} refer to experiments run with  the following settings w.r.t. Algorithm~\ref{alg:weightopt}: $n_0=10$, i.e., the initial best  vector of weights $W_{best}$  has been selected  amongst 10 random vectors and  $max_{iter}=50$, $\delta=10^{-3}$ i.e.,  termination of minimisation either after maximum 50 iterations or if the relative decrement of the minimised distance (LH or EMD) is less than $10^{-3}$.

Results in the 'Best estimator' columns refer instead to sWN models whose weights are obtained by either one amongst the six weight estimators introduced in~\cite{DBLP:conf/icpm/BurkeLW20}, namely: 
the frequency estimator (freq), the
activity-pair frequency estimators (lhpair, resp. rhpair), the 
mean-scaled activity-pair frequency estimator (pairscale), the  
fork distribution estimator (fork) and the 
alignment estimator  (align). Column 'name' indicates the name of the 'best estimator' (amongst the 6 ones) meaning the estimator that yield the sWN which has the best (i.e. smaller) distance from the stochastic language of the corresponding log, while columns 'LH' and 'rEMD' report about the LH, resp. rEMD distances measured on the sWN given by the best estimator.  

Columns marked 'tEMD' instead report the tEMD distance evaluated w.r.t. a probability mass threshold of 0.8 (as classically done e.g. in~\cite{DBLP:conf/icpm/BurkeLW20}). These results have been obtained by importing the resulting sWN models (i.e. with transitions weight obtained either by optimisation or given by Burke's estimators) into the ProM platform~\cite{10.1007/11494744_25} which amongst many functionalities supports the computation of tEMD  between a sWN and the corresponding event log. Entries marked 'TO' (as in Time Out) indicates cases in which we were not able to obtain the tEMD value as tEMD computation either crashed in ProM (unfortunately ProM seems to suffer of stability issues in some case) or was stopped as taking too long. 

Finally column 'RSD' refer to sWN models directly mined from the logs through the direct stochastic process  discovery  approach given in~\cite{5f0e8dd04572478fb450eefe4211d69f} whose implementation is included in ProM. 

By comparing, in Table~\ref{tab:exp1}, the  LH and EMD distances (in columns LH-opt, rEMD-opt) with those obtained with the estimators (column Best estimator)  we observe that for all considered logs, weight optimisation yields considerably reduced distances w.r.t. to the estimators' ones. For example for BPIC20\_dd, when using EMD optimisation we get a optimised rEMD distance of 0.02 as opposed to 0.93 obtained with the best (\emph{fork}) estimator  and, similarly, with LH optimisation we obtain a LH distance of 3.55 versus 25.15 which results from the best estimator.  % of in terms of  a considerably more conform sWN
%The 'LOGLH-opt' column of Table~\ref{tab:exp1} 

%displays the results of optimizing the log-likelihood with the 'L-BFGS-B' method, while the 'EMD-opt' column showcases the outcomes of optimizing the Earth Mover's Distance (EMD) with the 'Powell' method. Both experiments begin by selecting the best weight vector from a panel of 10 random vectors. Subsequently, the optimization process evaluates whether the relative improvement of the new vector is significantly better than the previous step, defined as an improvement greater than a threshold of $0.001$. The optimization terminates after $50$ iterations or upon convergence to a local minimum. The permissible range for weight values is between $0.001$ and $1$.

%%%%%%%% TABLE 2 %%%%%%%%%%
%\input{pic_latex/table2}
%%%%%%%%%%%%%%%%%%%%%%%%%%%
%%%%%%%% TABLE 3 %%%%%%%%%%
\begin{table}[]
    \centering
    \begin{tabular}{l p{1ex} l l p{1ex} l l l p{2ex} l l p{2ex} l l}
        \toprule 
        Event log & &  \multicolumn{2}{l}{log properties} & & \multicolumn{3}{l}{Computation time} & & \multicolumn{2}{l}{LH-opt} & & \multicolumn{2}{l}{rEMD-opt} \\
        \cmidrule(){3-4} \cmidrule(){6-8} \cmidrule(){10-11} \cmidrule(){13-14}
         & &  \#traces & \#act & & unfold & LH & rEMD & & $\sharp$iter & time & & $\sharp$iter & time \\
        \hline
        BPIC13\_c & & 183 & 4 & & 0.18 & 0.00029 & 0.096 & & 4 & 36.58 & & 8 & 603.12 \\
        BPIC13\_i & & 1511 & 4 & & 8.15 & 0.0046 & 8.06 & & 4 & 1913.23 & & 8 & 87086.48 \\
        BPIC13\_o & & 108 & 3 & & 0.072 & 0.00025 & 0.036 & & 8 & 35.18 & & 6 & 137.68 \\
        BPIC17\_ol & & 16 & 8 & & 0.00029 & 2.05e-05 & 0.00057 & & 12 & 0.11 & & 11 & 1.27 \\
        BPIC20\_dd & & 99 & 17 & & 0.19 & 0.00027 & 0.035 & & 14 & 296.32 & & 31 & 3946.85 \\
        BPIC20\_rfp & & 89 & 19 & & 0.48 & 0.0003 & 0.025 & & 7 & 572.85 & & 11 & 59819.73 \\
        Roadfines & & 231 & 11 & & 1.01 & 0.0014 & 0.16 & & 5 & 204.3 & & 4 & 1276.95 \\
        \bottomrule
    \end{tabular}
    \caption{Log's characteristics and corresponding computation time  (in seconds) for weight optimisation.}
    \label{tab:timeprop}
\end{table}

%%%%%%%%%%%%%%%%%%%%%%%%%%%

Table \ref{tab:timeprop} reports about relevant statistics of considered logs, as well as computation time to apply our optimisation framework on them.
%compu of the corresponding mined WN and resulting reachability graph. 
The  number of unique traces, 
respectively of activities of a log are given  in columns '$\sharp$traces', respectively '$\sharp$act'. % The number of place, respectively transitions of the mined WN are given in columns '$\sharp$pl, resp. '$\sharp$tr' while columns '$\sharp$state' and '$\sharp$tr' give the number of states, resp. edges of the corresponding RG. 
%a description of the data utilized in our study. The 'log properties' column delineates the dimensions of the objective language, while the 'net properties' column outlines the size of the control flow net mined with inductive miner to which stochastic properties are being added. Finally, the 'reachability graph' column offers insight into the complexity of the net, indicating the number of markings and possible traces it can reproduce, which serves as an indicator of the time required to unfold it and obtain its stochastic language.
%%%%%%%% TABLE 3 %%%%%%%%%%
%\input{pic_latex/table3}
%%%%%%%%%%%%%%%%%%%%%%%%%%%
 Column 'unfold' denotes the runtime for computing the sWN stochastic language via unfolding (Algorithm~\ref{alg:traceprob}), while 'LH', resp. 'rEMD' the runtime for computing the LH, resp. rEMD, distance for the obtained stochastic language returned by Algorithm~\ref{alg:traceprob}.Columns '$\sharp$iter' give the number of iterations executed during optimisation while 'time' the optimisation runtime (Algorithm~\ref{alg:weightopt}). We point out that the number of traces in the log affect   optimisation making LH optimisation preferable to rEMD as the cost for  rEMD distance  computation is much higher (e.g. BPIC13\_i). 
 
\begin{figure}
    \centering
    \scalebox{.92}{
    \begin{tabular}{cc}
         \pgfplotsset{every axis/.append style={
                    axis x line=bottom,    % put the x axis in the middle
                    axis y line=left,    % put the y axis in the middle
                    axis line style={->}, % arrows on the axis
                    xlabel={$x$},          % default put x on x-axis
                    ylabel={$y$},          % default put y on y-axis
                    ylabel style={rotate=-90, anchor=south, outer sep=50pt, xshift=40pt}, % rotate the ylabel and anchor it at the top
                    label style={font=\tiny},
                    tick label style={font=\tiny}  
                    }}

\begin{tikzpicture}
        \begin{axis}[
        legend style={nodes={scale=0.5, transform shape}},
        width=0.55\textwidth,
      height=0.4\textwidth,
            xlabel={$\sharp$ iterations},
            ylabel={Loglh},
            xmin=0, xmax=10,
            ymax=30,
            xtick={0,2,4,6,8,10},
            ymajorgrids=true,
            grid style=dashed,
            legend columns=2,
            ymode=log,
            %log ticks with fixed point,
            scaled y ticks=real:1e3
        ]
        
        \addplot[
            color=red,
            mark=*,
            ]
            coordinates {
            (0,12.987222893813243)(1,3.9826427722957227)(2,3.9432686793248553)(3,3.7550719118241442)(4,3.7272172450139425)(5,3.7271846393554644)(6,3.7253810423535554)(7,3.723120815277143)(8,3.720960277350923)(9,3.7205362303295657)(10,3.7181520304409563)
            };
        \addlegendentry{BPIC13\_c}

        \addplot[
            color=green,
            mark=square,
            ]
            coordinates {
            (0,15.687323586674227)(1,11.93131122785461)(2,10.451779913976692)(3,9.555427315230443)(4,9.364084808669315)(5,8.863361717121872)(6,8.720737525951668)(7,8.701508045375979)(8,8.69776809053902)(9,8.679449566258365)(10,8.675518468385052)
            };
        \addlegendentry{BPIC13\_i}

        \addplot[
            color=blue,
            mark=triangle,
            ]
            coordinates {
            (0,9.84234083739781)(1,6.616293638416069)(2,4.8104483269007865)(3,4.762217415342507)(4,4.3752651032561)(5,4.364807888333388)(6,4.193539990151345)(7,4.036071666849705)(8,3.9549941539096998)(9,3.954715415965181)(10,3.941643461562925)
            };
        \addlegendentry{BPIC13\_o}

        \addplot[
            color=orange,
            mark=diamond,
            ]
            coordinates {
            (0,4.6563219304941015)(1,4.268688577780792)(2,2.5852383647415222)(3,2.4973823608657466)(4,2.4481356394361846)(5,2.110388075641239)(6,1.9932887576651148)(7,1.9735374733043833)(8,1.9721207198546633)(9,1.9703806324569448)(10,1.9698399004086982)
            };
        \addlegendentry{BPIC17\_ol}

        \addplot[
            color=magenta,
            mark=x,
            ]
            coordinates {
            (0,15.111215394010676)(1,6.828910658963665)(2,6.515059910384583)(3,4.713418212581694)(4,4.20965591737279)(5,3.8569317592025127)(6,3.6704284786198467)(7,3.6488355964770585)(8,3.600842433570134)(9,3.5883451484646374)(10,3.5261446118981734)
            };
        \addlegendentry{BPIC20\_dd}

        \addplot[
            color=cyan,
            mark=+,
            ]
            coordinates {
            (0,29.02092943796397)(1,7.347155918377557)(2,6.765141448678697)(3,6.7437913743031395)(4,6.405556069266726)(5,5.756575388552254)(6,5.3958169386142085)(7,5.390405818405931)(8,5.208305705425694)(9,5.1976610210845084)(10,4.854948761300046)
            };
        \addlegendentry{BPIC20\_rfp}

        \addplot[
            color=black,
            mark=o,
            ]
            coordinates {
            (0,7.99355706658853)(1,4.524329316180952)(2,4.155175871805369)(3,3.7303945050675664)(4,3.724861069233226)(5,3.5468419091554484)(6,3.546227681003157)(7,3.4601114068086543)(8,3.421883155290012)(9,3.4139885063275894)(10,3.3432779332463247)
            };
        \addlegendentry{Roadfines}
            
        \end{axis}
    \end{tikzpicture} & \pgfplotsset{
    every axis/.append style={
        axis x line=bottom,    % put the x axis at the bottom
        axis y line=left,      % put the y axis on the left side
        axis line style={->},  % arrows on the axis
        xlabel={$x$},          % default put x on x-axis
        ylabel={$y$},          % default put y on y-axis
        ylabel style={rotate=-90, anchor=south, outer sep=50pt, xshift=40pt}, % rotate the ylabel and anchor it at the top
        label style={font=\tiny},
        tick label style={font=\tiny}
    }
}

\begin{tikzpicture}

    \begin{axis}[
        legend style={nodes={scale=0.5, transform shape}},
        width=0.55\textwidth,
        height=0.4\textwidth,
        xlabel={$\sharp$ iterations},
        ylabel={EMD},
        xmin=0, xmax=10,
        ymin=0, ymax=1,
        xtick={0,2,4,6,8,10},
        ytick={0,0.2,0.4,0.6,0.8,1},
        ymajorgrids=true,
        grid style=dashed,
        legend columns=2
    ]
        
    \addplot[
        color=red,
        mark=*,
        ]
        coordinates {
        (0,0.6181407053355435)(1,0.12371600498308592)(2,0.09362848342668637)(3,0.08805515847771452)(4,0.08087586769011453)(5,0.07658084508537165)(6,0.07427930052154426)(7,0.07340899188537175)(8,0.07285407980588594)(9,0.07231173697851476)(10,0.07175061366537153)
        };
    \addlegendentry{BPIC13\_c}

    \addplot[
        color=green,
        mark=square,
        ]
        coordinates {
        (0,0.6479456575250001)(1,0.4964539141660018)(2,0.20738041482900518)(3,0.2024645568210013)(4,0.1989171409290024)(5,0.1926582955460052)(6,0.1852736015830032)(7,0.1820337753310049)(8,0.1818916441770011)(9,0.1818916441770011)(10,0.1818916441770011)
        };
    \addlegendentry{BPIC13\_i}

    \addplot[
        color=blue,
        mark=triangle,
        ]
        coordinates {
        (0,0.25798807606000024)(1,0.2275098062960003)(2,0.16938798583300013)(3,0.10251803453900067)(4,0.08778506469400024)(5,0.07771390531200023)(6,0.07336604942900045)(7,0.07322394249400045)(8,0.07317527496100071)(9,0.07311735292999999)(10,0.07307328274200055)
        };
    \addlegendentry{BPIC13\_o}

    \addplot[
        color=orange,
        mark=diamond,
        ]
        coordinates {
        (0,0.2362842365142)(1,0.11051233744559999)(2,0.09362943733020013)(3,0.0934765708512)(4,0.093469854198)(5,0.0934614875496002)(6,0.09344838757080014)(7,0.09342500431800006)(8,0.093417637647)(9,0.09338968773840016)(10,0.0933166381074)
        };
    \addlegendentry{BPIC17\_ol}

    \addplot[
        color=magenta,
        mark=x,
        ]
        coordinates {
        (0,0.8301273000000003)(1,0.116920051286)(2,0.05506652855000071)(3,0.05001243053300094)(4,0.042202929207000466)(5,0.024979401757000554)(6,0.022139519077000226)(7,0.020812439843000113)(8,0.019809220893000558)(9,0.019053560734000455)(10,0.01849435107)
        };
    \addlegendentry{BPIC20\_dd}

    \addplot[
        color=cyan,
        mark=+,
        ]
        coordinates {
        (0,0.9893190000000005)(1,0.5744460312950002)(2,0.5664705877750004)(3,0.5647592891270004)(4,0.5634047149020002)(5,0.5560813021490001)(6,0.5530807720540004)(7,0.5515229970500001)(8,0.5496944551110001)(9,0.5475002656870002)(10,0.5443563711830002)
        };
    \addlegendentry{BPIC20\_rfp}

    \addplot[
        color=black,
        mark=o,
        ]
        coordinates {
        (0,0.3279948621181)(1,0.09922119946795)(2,0.09896491136925045)(3,0.09890040504970045)(4,0.09882867718165157)(5,0.0987466628276001)(6,0.09868688297845031)(7,0.0986342582348004)(8,0.09846928847570093)(9,0.0984304236109003)(10,0.09836061903400094)
        };
    \addlegendentry{Roadfines}
            
    \end{axis}
\end{tikzpicture} \\
    \end{tabular}
    }
    \caption{LH and EMD convergence with iterative optimisation.}
    \label{fig:convergence}
\end{figure}
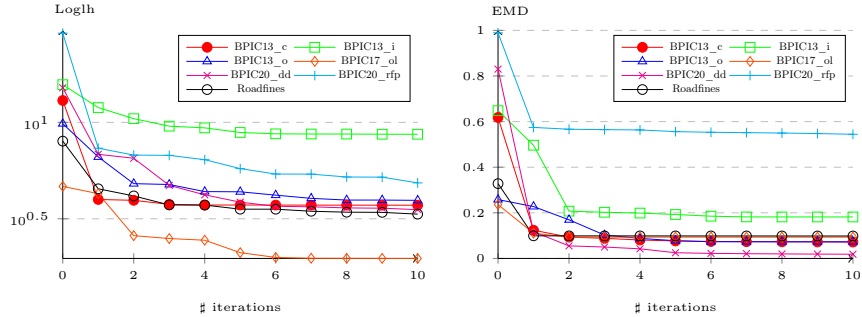

Finally, Figure~\ref{fig:convergence} depicts results concerning the convergence of distances between the net's and the log's stochastic languages  measured during the iterative steps of minimisation and w.r.t. LH distance (left) and rEMD distance (right). 

%Table \ref{tab:exp1} presents the comparative results among our method, the best weights estimators, and the outcomes of RSD. The 'LOGLH-opt' column displays the results of optimizing the log-likelihood with the 'L-BFGS-B' method~\cite{10.1145/279232.279236}, while the 'EMD-opt' column showcases the outcomes of optimizing the Earth Mover's Distance (EMD) with the 'Powell' method. Both experiments begin by selecting the best weight vector from a panel of 10 random vectors. Subsequently, the optimization process evaluates whether the relative improvement of the new vector is significantly better than the previous step, defined as an improvement greater than a threshold of $0.001$. The optimization terminates after $50$ iterations or upon convergence to a local minimum. The permissible range for weight values is between $0.001$ and $1$.

\section{Conclusion}
\label{sec:conclusion}

We have tackled the problem of obtaining a stochastic model that is capable of reproducing the likelihood of behaviors observed on a real-life system and stored in an event log   in terms of unique traces equipped with frequency. 
This poses a relevant issue as classical process mining approaches ignore trace frequency hence  yield models incapable of accounting for probabilistic behaviour exhibited by a system.
Existing stochastic discovery approaches introduced Petri net's weight estimators~\cite{DBLP:conf/icpm/BurkeLW20} whose values are obtained by combining  summary statistics computed on the log and on structurally related transitions of the net. Although computationally light these estimators have a limited ability to accurately reproduce the stochastic character of the observed system. In this paper we have introduced a novel weight estimation framework which, relying on an optimisation scheme, search for optimal weights, yielding a stochastic Petri net that closely reproduce the log's stochastic language. The framework assesses the net's stochastic language by unfolding of the corresponding RG and applies minimisation w.r.t. to a stochastic language distance metric. More specifically we let the user opt between either  a \emph{likelihood} based or an earth mover's based distance, which can be chosen to drive optimisation also in function of the complexity of the log (with LH optimisation being computationally lighter to be prefered for large logs). 
We have demonstrated the optimisation framework w.r.t. different real-life logs and showed that it yields considerably more accurate models than alternative approaches.

There are several aspects worth considering as future developments. These include improving the complexity of the evaluation of the net's stochastic language, currently the bottleneck in our framework, as well as considering other kinds of  language's distance such as the  entropy related ones which are more suitable to deal with infinite languages~\cite{Polyvyanyy2020AnER}.

%Utilizing the optimizer across diverse real-life datasets of varying sizes, model complexities, and patterns enables us to generate high-quality Generalized Stochastic Petri Nets (GSPNs). This approach ensures that the behavior captured by the mined model closely aligns with that described in the log data, encompassing both control flow and stochastic aspects. 

%\begin{credits}
%\subsubsection{\ackname} A bold run-in heading in small font size at the end of the paper is used for general acknowledgments, for example: This study was funded by X (grant number Y).

%\subsubsection{\discintname}
%The authors have no competing interests to declare that are relevant to the content of this article.
%\end{credits}

\bibliographystyle{plain} 
\bibliography{biblio}

%\appendix
%\input{sections/6_appendix}

%\input{sections/6_garbage}

\end{document}